\documentclass{ws-procs9x6}
\newcommand{\be}{\begin{equation}}
\newcommand{\ee}{\end{equation}}
\newcommand{\bea}{\begin{eqnarray}}
\newcommand{\eea}{\end{eqnarray}}
\newcommand{\nn}{\nonumber}
\begin{document}

\title{On asymmetric Chern-Simons number diffusion}

\author{Bert-Jan Nauta}

\address{Institute for Theoretical Physics, University of Amsterdam\\
Valckenierstraat 65, 1018XE, Amsterdam, The Netherlands, \\
E-mail: nauta@science.uva.nl}

\maketitle

\abstracts{ We show that CP-violation can lead to an asymmetric
diffusion of the Chern-Simons number in thermal equilibrium. This
asymmetry leads to a linearly growing expectation value of the
third power of the Chern-Simons number. In the long-time limit all
expectation values of powers of Chern-Simon numbers are determined
by their appropriate disconnected parts. }

\section{Chern-Simons number diffusion}
At high temperatures the sphaleron rate determines the rate of
Chern-Simons number changing transitions. Since a change in the
Chern-Simons number is accompanied by a (3 times larger) change in
the baryon number (in the standard model), the sphaleron rate is
an important ingredient of baryogenesis scenarios. The sphaleron
rate may be obtained by ignoring the back reaction of the
generated baryons and leptons. This back reaction may then later
be included through a linear response analysis. Without (the back
reaction of) baryons and leptons the potential is periodic (as for
instance in pure gauge-Higgs theory). When subsequent transitions
are uncorrelated or the correlation dies away sufficiently fast,
the process is Markovian and the Chern-Simons number diffuses. The
diffusion process may be
specified by \bea \langle N_{\rm CS}(t)-N_{\rm CS}(0)\rangle &=& 0,\label{1stpower}\\
 \langle\left[N_{\rm CS}(t)-N_{\rm CS}(0)\right]^2\rangle &=& 2V\Gamma_{\rm sph}
 t,\label{2ndpower}
\label{CSdiff}\eea with \(\Gamma_{\rm sph}\) the sphaleron rate
and \(V\) the volume. See e.g. the review \cite{review}.

CP-violation is one of three of Sakharov's requirements for
baryogenesis. This is our reason to study the effect of
CP-violation on Chern-Simons number diffusion. Since the
Chern-Simons number itself is CP-odd, CP-violation may cause odd
powers of the Chern-Simons number to take non-zero values. The
question that we want to answer is: what is the possible behavior
of the expectation values \(\langle \left[N_{\rm CS}(t)-N_{\rm
CS}(0)\right]^{2n+1}\rangle\) for \(n=0,1,..\) in thermal
equilibrium when CP is violated.

\section{CPT invariance and thermal equilibrium}
In general, the expectation values of CPT-odd quantities, such as
the Chern-Simons number, vanish in thermal equilibrium. Before we
will show that this does not apply to odd powers of differences of
Chern-Simons numbers, we review the argument.

Consider  the expectation value of a CPT-odd operator, \( B\), for
instance the baryon number. Since in thermal equilibrium the
(thermal) distribution function  is time-independent, the thermal
average of \(B(t)\) may be taken at time \(t\)
 \be \langle B(t)\rangle = \langle B \label{Blocal}\rangle.
\label{Bth}\ee The thermal average is defined by \(\langle B
\rangle = \mbox{Tr}\left\{ B e^{-\beta H} \right\},\) where
\(\beta\) is the inverse temperature and \(H\) is the Hamiltonian.

Since the operator \(B\) is CPT-odd and the Hamiltonian is
CPT-even
 \be \langle B \rangle
\stackrel{\mbox{CPT}}{=}-\langle B \rangle \;\;\;\Rightarrow\;\;\;
\langle B \rangle =0.\label{Bvanishes}\ee
 see e.g. \cite{bernreuther}. In
general, the thermal expectation value of CPT-odd operators
vanishes.

Basically there are two conditions for this argument to work: the
CPT-odd operator must be local in time, and the thermal average
must exist.

\section{ Why \(\langle \left[N_{\rm CS}(t)-N_{\rm
CS}(0)\right]^{2n+1}\rangle\) can be non-zero}

 If we try to repeat the reasoning that leads to the conclusion
that CPT-odd quantities vanish in thermal equilibrium (from
(\ref{Bth}) to (\ref{Bvanishes})) for odd powers of differences in
Chern-Simons numbers at different times, we run into the following
problem. If we directly apply a CPT-transformation we arrive at
\be \langle \left[N_{\rm CS}(t)-N_{\rm
CS}(0)\right]^{2n+1}\rangle\stackrel{\mbox{CPT}}{=} -\langle
\left[N_{\rm CS}(-t)-N_{\rm CS}(0)\right]^{2n+1}\rangle\ee A time
translation on the r.h.s. shows that this is a trivial identity.
The argument leading to (\ref{Bvanishes}) only works for CPT-odd
operators that are local in time, so that the thermal average can
be taken at the time the operator is evaluated, as in
(\ref{Blocal}). With this in mind, we might try to apply the
reasoning to \(\langle N_{\rm CS}^{2n+1}(t)\rangle\) and we would
find it vanishes. However only differences of Chern-Simons numbers
are gauge-invariant. The thermal average \(\langle N_{\rm
CS}^{2n+1}(t)\rangle\) does not exist, and the argument leading to
(\ref{Bvanishes}) cannot be applied in this case. It is important
to realize that there is no way around this: only differences
\(N_{\rm CS}(t_1)-N_{\rm CS}(t_2)\) are gauge invariant and
physical, but a thermal average of powers of these differences can
be taken only at a single time \(t_1\), \(t_2\), or some other
time. As a result these quantities may depend on \(t_1-t_2\) and
do not need to vanish.

Hence, we conclude that the standard argument that the thermal
average of CPT-odd quantities vanishes based on CPT-invariance is
not applicable to odd powers of the difference of Chern-Simons
numbers.

The combination of CPT-invariance and thermal equilibrium still
make gauge-invariant CPT-odd quantities that depend on a single
time vanish, such as \be \langle\left\{N_{\rm CS}(t) -
\mbox{round}\left[N_{\rm CS}(t)\right]\right\}^{2n+1}\rangle=0,
\label{oddvanish}\ee with round the function that maps a real
number to its nearest integer. Equations as (\ref{oddvanish}) pose
restrictions on the distribution function of the Chern-Simons
number. But it poses no restrictions on the expectation values we
are interested in, expectation values of powers of differences of
the Chern-Simons number.

But, there is an additional important restriction, that follows
from \(\langle \frac{\mbox{d}}{\mbox{dt}}\left[N_{\rm
CS}(t)-N_{\rm CS}(0)\right]\rangle =0,\) since the thermal average
of a time derivative of a quantity that is local in time vanishes.
This gives \be \langle \left[N_{\rm CS}(t)-N_{\rm
CS}(0)\right]\rangle =0.\label{singlepowzero}\ee For expectation
values of higher powers there is not such a restriction \(
\langle\frac{\mbox{d}}{\mbox{dt}}\left[N_{\rm CS}(t)-N_{\rm
CS}(0)\right]^{m}\rangle \not =0,\; m=2,3,4,...\) as is well-known
for \(m=2\).

Hence, we may summarize that on general grounds only the
expectation value of a single power of a difference in
Chern-Simons numbers vanishes in thermal equilibrium
(\ref{singlepowzero}), but that generically \be\langle
\left[N_{\rm CS}(t)-N_{\rm CS}(0)\right]^{m}\rangle \not=0,
\label{oddhigherpownonzero}\ee for all, including odd, \(m\geq
2\). Here it should be understood that (\ref{oddhigherpownonzero})
means that it is {\em possible} that these expectation values are
non-zero. Of course for the expectation values of odd powers to
actually obtain a non-zero value, at least CP has to violated.
(Formally, this can be shown by the same argument as leading to
(\ref{Bvanishes}), but with a CP-transformation instead of a
CPT-transformation.)

\section{Asymmetric diffusion}
We have argued that when CP is violated there is no reason why
\(\langle\left[N_{\rm CS}(t)-N_{\rm CS}(0)\right]^{2n+1}\rangle\)
with \(n=1,2,3...\) should be equal to zero in thermal
equilibrium.

For Chern-Simons number diffusion to develop asymmetrically, an
asymmetry must exist in sphaleron transitions. We have
investigated such an asymmetry by considering the effect of
CP-violation on the motion of configurations at the sphaleron
barrier\cite{nauta}. The CP-violation was added by including
effective CP-odd operators to the action. The lowest order
operator, that is mostly considered, is the dimension-six
operator, \(\phi^{\dagger}\phi \tilde{F}_{\mu\nu}^a F^{\mu\nu
a}\). In out-of-equilibrium situation this operator can generate
an effective chemical potential for baryons, and therefore it
introduces a bias in the sphaleron transitions\cite{dimsix}.
However, we found that, within the chosen approach, in thermal
equilibrium the dimension-six operator does not lead to any
asymmetry\cite{arrizabalaga}. But we also found that CP-odd
dimension-eight operators induce an asymmetry in the average
velocity with which a configuration crosses the sphaleron
barrier\cite{arrizabalaga}. Unfortunately, we have not succeeded
to relate this asymmetry in the velocity to an asymmetry in the
distribution of the Chern-Simons number after several sphaleron
transitions. Nevertheless, one may assume the the distribution to
develop an asymmetry, at least, there is absolutely no reason why
it should not.

Next we determine the large-time behavior of the expectation
values of odd powers of Chern-Simons number differences. We start
with the third power. It can be shown that in the long-time limit
\(\langle \left[N_{\rm CS}(t)-N_{\rm CS}(0)\right]^{3}\rangle\) in
thermal equilibrium either goes to a constant or grows linearly in
time (for a one-dimensional system this was shown in
\cite{arrizabalaga}).

 When the sphaleron barrier is large,
subsequent transitions are practically uncorrelated. In that case,
an asymmetry in the sphaleron transitions will lead to a linearly
growing value in time for \(\langle \left[N_{\rm CS}(t)-N_{\rm
CS}(0)\right]^{3}\rangle\)\cite{arrizabalaga}. Hence in a theory
with CP-violation, an additional equation is required to describe
the (asymmetric) diffusion of the Chern-Simons number
 \be
\langle \left[N_{\rm CS}(t)-N_{\rm CS}(0)\right]^{3}\rangle =
\delta_{\rm CP}V\Gamma_{\rm sph}t, \label{3rdpower}\ee with
\(\delta_{\rm CP}\) a dimensionless parameter proportional to the
coefficients of the CP-odd operators in the theory. Note that the
assumption that the sphaleron barrier is large implies that we
have only argued (\ref{3rdpower}) to hold in the broken phase. We
would be surprised if in the symmetric phase (\ref{3rdpower}) is
incorrect, but we do not have any physical arguments for this.

Equation (\ref{3rdpower}) can be tested in a one-dimensional
model. All arguments leading to (\ref{3rdpower}) apply also to a
model of a particle with (CP-odd) coordinate $x$ with a Lagrangian
that is not invariant under \( x \rightarrow -x\), but that is
invariant under the combined reflections of \(t \rightarrow -t\)
and \( x \rightarrow -x\) (T and CP), and has a periodic potential
(with a barrier between different vacua). Such a one-dimensional
model\cite{arrizabalaga} was coupled to a thermal bath and the
diffusion was studied, the result showed indeed a linearly growing
value for \(\langle x^3\rangle\) proportional to the diffusion
constant of \(\langle x^2\rangle\).

To relate the constant \(\delta_{\rm CP}\) in (\ref{3rdpower}) to
the coefficient of certain CP-odd operator seems to be very
difficult analytically, since sphaleron transitions require
non-perturbatively large fields. Also numerically this may be a
problem, since in 3+1D the classical theory is divergent and this
situation may worsen when non-renormalizable (effective) CP-odd
operators are added to a theory. We are performing a numerical
study in 1+1D.
Our
preliminary results confirm (\ref{3rdpower}).

Expectation values of higher odd powers are expected to be
dominated by their disconnected parts in the large-time
limit\cite{arrizabalaga} \bea \langle \left[N_{\rm CS}(t)-N_{\rm
CS}(0)\right]^{2n+1}\rangle & \approx
&\frac{(2n+1)!}{2^{n-1}3!(n-1)!}  \langle \left[N_{\rm
CS}(t)-N_{\rm CS}(0)\right]^3\rangle \times \nn\\&& [\langle
\left[N_{\rm CS}(t)-N_{\rm CS}(0)\right]^2\rangle]^{n-1}. \eea In
this way all correlation function are determined by the set of
equations (\ref{1stpower}), (\ref{2ndpower}), and
(\ref{3rdpower}).

\section{Conclusion and outlook}
We have shown that CP-violation can induce an asymmetry in the
distribution function of the Chern-Simons number. We have argued
that in the broken phase in the presence of CP-violation
Chern-Simons number diffusion is determined by the set of
equations (\ref{1stpower}), (\ref{2ndpower}), and
(\ref{3rdpower}).

The study of asymmetric Chern-Simons number diffusion is motivated
by the the question how the baryon asymmetry in the universe was
generated. It has been argued that asymmetric diffusion may result
in a non-zero expectation value of the baryon number in a
time-dependent effective potential \cite{arrizabalaga}.


\begin{thebibliography}{0}
\bibitem{review} V. A. Rubakov and M. E. Shaposhnikov, {\it Phys. Usp.} {\bf
39} 461 (1996).
\bibitem{bernreuther} W. Bernreuther, {\it Lect. Notes Phys.} {\bf 591}
237 (2002).
\bibitem{nauta} B. J. Nauta, {\it Phys. Lett.} {\bf B478} 275 (2000).
\bibitem{dimsix} J. Garcia-Bellido, D. Y. Grigoriev, A. Kusenko, and M.E. Shaposhnikov, {\it Phys. Rev.} {\bf D60} 123504
(1999);
J. Smit and A. Tranberg, hep-ph/0211243, and
references therein.
\bibitem{arrizabalaga} B. J. Nauta and A. Arrizabalaga, {\it Nucl. Phys.} {\bf B635},
255 (2002).



\end{thebibliography}
\end{document}